\documentclass{emulateapj}

\usepackage{units}
\usepackage{amsmath}
\usepackage{graphicx}

\newcommand{\dd}[1]{\mathrm{d}#1}
\newcommand{\msun}{\mathrm{M}_\odot}

\defcitealias{oni02}{OMK02}

\shorttitle{Star Formation in the Taurus filament L\,1495}
\shortauthors{Schmalzl et al.\ (2010)}

\begin{document}

\title{Star formation in the Taurus filament L\,1495: From Dense Cores to Stars
\footnote{Based on observations collected at the
German-Spanish Astronomical Center, Calar Alto, jointly operated by the
Max-Planck-Institut f\"ur Astronomie Heidelberg and the
Instituto de Astrof\'isica de Andaluc\'ia (CSIC)}}

\author{Markus Schmalzl, Jouni Kainulainen, Ralf Launhardt, Thomas Henning}
\affil{Max-Planck-Institut f\"ur Astronomie, K\"onigstuhl 17, 69117 Heidelberg, Germany}
\email{schmalzl@mpia.de}
\author{Sascha P.\ Quanz}
\affil{Institute for Astronomy - Swiss Federal Institute of Technology (ETH), Wolfgang-Pauli-Strasse 27, 8093~Zurich, Switzerland}
\author{Jo\~ao Alves}
\affil{University of Vienna, T\"urkenschanzstrasse 17, 1180 Vienna, Austria}
\author{Carlos G.\ Rom\'an-Z\'u\~niga}
\affil{Centro Astron\'omico Hispano Alem\'an/Instituto de Astrof\'isica de Andaluc\'ia (IAA-CSIC). Glorieta de la Astronom\'ia, S/N, Granada 18008, Spain}
\author{Alyssa A.\ Goodman, Jaime E.\ Pineda}
\affil{Harvard-Smithsonian Center for Astrophysics, 60 Garden Street, MS 42, Cambridge, MA 02138, USA}

\begin{abstract}
We present a study of dense structures in the L\,1495 filament in
the Taurus Molecular Cloud and examine its star-forming properties. In
particular we construct a dust extinction map of the filament using
deep near-infrared observations, exposing its small-scale structure in
unprecedented detail. The filament shows highly fragmented substructures and a
high mass-per-length value of
$M_\mathrm{line}=\unit[17]{\msun\,pc^{-1}}$,
reflecting star-forming potential in all parts of it. However, a part of
the filament, namely B\,211, is remarkably devoid of young stellar objects. We
argue that in this region the initial filament collapse and fragmentation is
still taking place and star formation is yet to occur. In the star-forming
part of the filament, we identify 39 cores with masses from
$0.4\ldots\unit[10]{\msun}$ and preferred separations in agreement with the local
Jeans length. Most of these cores exceed the
Bonnor-Ebert critical mass, and are therefore likely to collapse and form stars.
The Dense Core Mass Function follows a power law with exponent
$\Gamma=1.2\pm0.2$, a form commonly observed in star-forming regions.
\end{abstract}



\section{Introduction}

\begin{figure*}[tb]
	\begin{center}
	\includegraphics[height=7in,angle=-90]{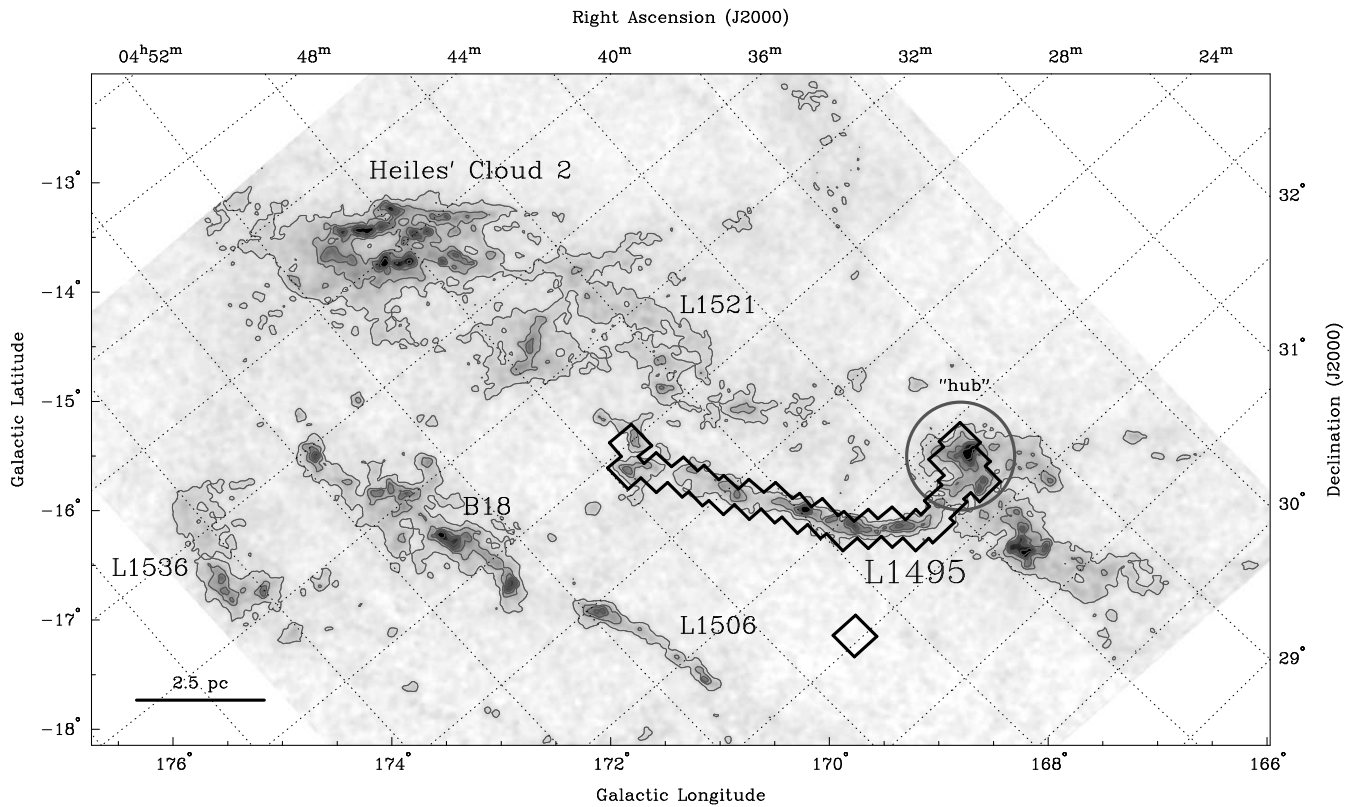}
	\caption{\label{fig01} Extinction map of the Taurus
	Molecular Cloud \citep{kai09} derived from 2MASS data.
	Contours are plotted in steps of $A_\mathrm{V}=\unit[2]{mag}$.
	The circle marks the center of the {\em hub-filament} system in L\,1495
	\citep{mye09} with three filaments emanating towards the east,
	south-west and west. The thick solid lines mark the region of our Omega2000
	observations in L\,1495, and the control field to the south of it.
	}
	\end{center}
\end{figure*}

Filaments appear to be common structural features in both
quiescent and star-forming molecular clouds.
In particular, recent observations of molecular clouds
with {\em Herschel} \citep{pil10} have signified the role of filaments as a
momentous, perhaps even dominant mode of star formation
\citep[e.g.][]{hen10,mol10,men10}. However, their formation and small scale
structure is still not well understood and under debate
\citep[see e.g.][and references therein]{mye09}.
\citet{sch79} noted that filaments tend to fragment into equally
spaced condensations of sizes $\sim\unit[10^0]{pc}$. These clumps, in turn,
consist of even smaller, gravitationally bound entities, namely cores, of sizes
$\sim\unit[10^{-1}]{pc}$, which is of the order of the local Jeans length.
These cores are then believed to be the direct precursors of stars and binary
systems \citep[see review by][]{ber07}.

Such a hierarchical structure can also be found in the Taurus Molecular Cloud
(Fig.\,\ref{fig01}),
which is one of the closest star-forming regions at a distance of only
$\unit[137\pm10]{pc}$ \citep{tor07}, and therefore an excellent testbed to 
study the small scale structure of filaments. Large scale surveys in CO
\citep{miz95,oni98,nar08} and mid-infrared dust emission \citep{reb10},
alongside with dust extinction maps \citep{dob05,kai09,lom10} have built a
detailed view of a complex network of filaments in Taurus,
which are known to be the birth sites of stars
\citep[see reviews about star formation in Taurus by e.g.,][]{pal02,ken08}.
These filament are composed of clumps \citep{bar27b,lyn62,lee99} with
sizes of $\sim\unit[1]{pc}$, which host a plethora of cores
with typical sizes of $\sim\unit[0.1]{pc}$. These 
have been detected e.g.\ via molecular line observations tracing cold and dense
gas with volume densities of $n_\mathrm{H_2}\gtrsim\unit[10^5]{cm^{-3}}$,
such as NH$_3$ \citep{jij99} and H$^{13}$CO$^+$
\citep[hereafter \citetalias{oni02}]{oni02}. Further down in this
hierarchical structure, from dust continuum observations
\citet{sad10} found a population of pre- and
protostellar cores, supposedly representing the scale of direct
core-protostar connection. The Taurus Molecular Cloud is known to host
$\gtrsim250$ Young Stellar Objects \citep[YSOs, ][]{reb10}, and their
distribution shows a strong correlation to the filamentary structures,
i.e.\ their birth sites \citep{har02,ken08}.

To further understand the connection between filamentary structures and star
formation, it is essential to characterize the filament in high spatial
resolution over a wide dynamic range. Molecular line species can be
used as probes for column densities, but typically rather exhibit a narrow
dynamical range. Larger dynamical range
can be achieved by thermal dust emission observations, which are used to trace
intermediate to high column densities, and dust extinction mapping, which
is more sensitive at low to intermediate column density range
\citep[see e.g.][for a comparison of different methods]{pin08,goo09,vas09}.
The aforementioned observations allow to determine the structure of the
low density clumps on one hand, and the core population detected by dense
gas tracers on the other hand. However, to make a direct connection between
the diffuse envelope structure and the denser cores, one preferably needs
a single tracer with a uniform calibration, which covers both these regimes.

Therefore, the aim of this work
is to create a column density map of 
L\,1495 \citep{lyn62} with unprecedented dynamical range and
resolution with the aid of near-infrared (NIR)
dust extinction mapping. NIR colors of stars
in the background of clouds provide hundreds to thousands of pencil beam
measurements of reddening, which can be smoothed out to construct maps
of column density along a molecular cloud. This is the basis of
NIR excess technique {\sc nice} and its derivatives
\citep{lad94,lom01,fos08,lom09}, which obtain
extinction measurements by correlating the observations from the
science field with a nearby control field. The color excess with respect
to this control field is then completely attributed to be caused by
the molecular cloud. These methods typically probe extinctions
$A_\mathrm{V}\sim0.5\ldots\unit[50]{mag}$ \citep[e.g.][]{lom01,kai06,kai07,rom09},
allowing us to examine both the detailed fragmentation of the filament into
dense cores, and the envelopes surrounding them. Thus, the observations
presented in this paper will provide the most complete census of the small
scale structures in the Taurus filament so far, and shed more light on the
link between filaments and star-forming cores.

The paper is organized as follows. In \S\ref{s-obs} we give an
overview of the observations and data reduction. In \S\ref{s-results} we present
our NIR extinction map of unprecedented depth and resolution, and derive the
dense core population, which is discussed in section \S\ref{s-discussion}.
Finally, the conclusions and summary follow in \S\ref{s-summary}.


\section{Observations and Data Reduction}
\label{s-obs}

We carried out NIR observations of the Taurus filament L\,1495 with the
Omega2000 camera at the Calar~Alto~3.5m telescope.
The camera is equipped with a HAWAII2 HgCdTe detector
that offers 2048$\times$2048 pixels with a pixel scale of
$\unit[0.45]{\arcsec}$ and a field of view of
$\unit[15.4\times15.4]{\arcmin}$. Observations were carried
out during two observing periods from October~2004
to December~2005 \citep{qua10} and January to March~2009. In total,
21~science fields and 1~control field were observed in
$J$, $H$, and $K_\mathrm{s}$ bands
($\unit[1.2]{\mu m}$, $\unit[1.6]{\mu m}$, and
$\unit[2.2]{\mu m}$, respectively). The locations of the
observed frames are shown in Fig.\,\ref{fig01}.

\begin{figure}[tb]
	\begin{center}
	\includegraphics[height=3.3in,angle=-90]{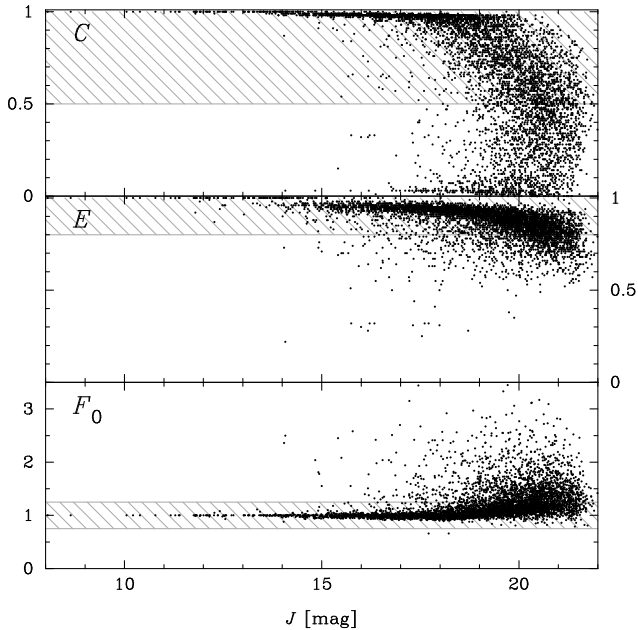}
	\caption{\label{fig02} Our star-galaxy separation is based on the
	stellarity index $C$, elongation $E$ and its normalized FWHM $F_0$. The
	hatched areas indicate the regions populated by stars. Due to
	illustrative reasons, only a small part of our final catalog is shown.}
	\end{center}
\end{figure}

\begin{figure}[tb]
	\begin{center}
	\includegraphics[height=3.3in,angle=-90]{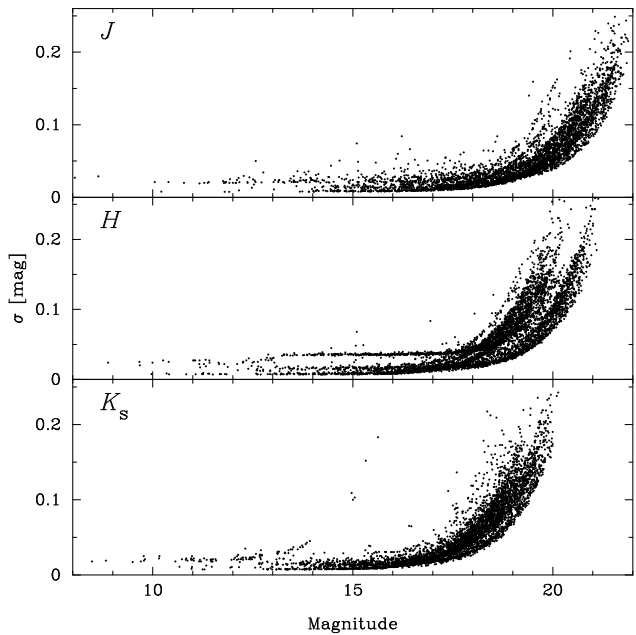}
	\caption{\label{fig03} Photometric uncertainties for our sources in
	the different filters. They contain contributions from the photometric
	error and the uncertainty
	of their individual photometric zero-point. This figure clearly shows the
	contributions from different frames, which differ in terms of sensitivity due
	to varying observing conditions. Due to
	illustrative reasons, only a small part of our final catalog is shown.}
	\end{center}
\end{figure}

The observations of a single Omega2000 field consisted of 30 dithered exposures,
each having an integration time of $\unit[60]{s}$. This resulted
in total exposure times of 30~minutes per filter per field.
We performed a standard NIR
data reduction with the Omega2000 pipeline \citep{fas03}, including flatfield,
dark, and bad pixel corrections, sky subtraction, and image registering.
The world coordinate system (WCS) was appended to the frames
with \verb+koords+ \citep{koords} by matching
the positions of at least 10~stars per frame to matching stars from the corresponding
2MASS image. This resulted to the rms accuracy of $\unit[0.1]{px}$
(or $\unit[0.045]{\arcsec}$) in all frames.

In addition to stars, a significant amount of background galaxies was expected
to be detected in the frames. In order to disentangle these two types of
sources from each other, we performed photometry using
{\sc SExtractor} \citep{sextractor}, which provides an automatic 
source classification via the so-called {\em stellarity index}
(\verb+CLASS_STAR+). In particular, photometry was
run in a fully automated two-pass mode. First, {\sc SExtractor} was run for
the brightest stars only, in order to get an accurate estimate of the
seeing, which is a crucial input parameter to get
reliable measures of the stellarity index, and furthermore determined
our choice of aperture size. Second, the actual photometry was performed,
resulting in instrumental magnitude and uncertainty,
position, the stellarity index $C$, elongation $E$ (the inverse of
the aspect ratio) and the parameter $F_0$, which is a
source's FWHM normalized by the seeing.

Each source was classified as either star or galaxy following
the classification scheme from \citet{can05}. In this scheme, a source
is regarded as a {\em star} if it fulfills two out of the three following
conditions:
\begin{itemize}
	\setlength{\itemsep}{-0.2ex}
	\item $C>0.5$
	\item $E>0.8$
	\item $0.75<F_0<1.25$
\end{itemize}
which are indicated as hatched regions in in Fig.\,\ref{fig02}.

The photometric calibration was done for each frame separately by matching
sources within the frame with stars from the 2MASS catalog, and determining
photometric zero-points and color corrections using those stars.
These catalogs in $J$, $H$ and $K_\mathrm{s}$ were then merged, and sources
with a detection in only one filter were rejected. Saturated stars
were replaced by their 2MASS counterparts. The final catalog
of L\,1495 contains $\sim$33,000 sources, of which 63\% are classified as
stars, according to our classification scheme. Photometric uncertainties for an
excerpt from this catalog are shown in Fig.\,\ref{fig03}. The median
magnitudes for which $\sigma=\unit[0.1]{mag}$ are at $J=\unit[20.6]{mag}$,
$H=\unit[19.4]{mag}$ and $K_\mathrm{s}=\unit[18.8]{mag}$, which is on average
$\unit[4]{mag}$ deeper than 2MASS, and emphasizes the depth of our
observations.

\section{Results}
\label{s-results}

\subsection{Dust Extinction in the Taurus Filament}
\label{s-gnicer}

\begin{figure}[tb]
	\begin{center}
	\includegraphics[height=3.3in,angle=-90]{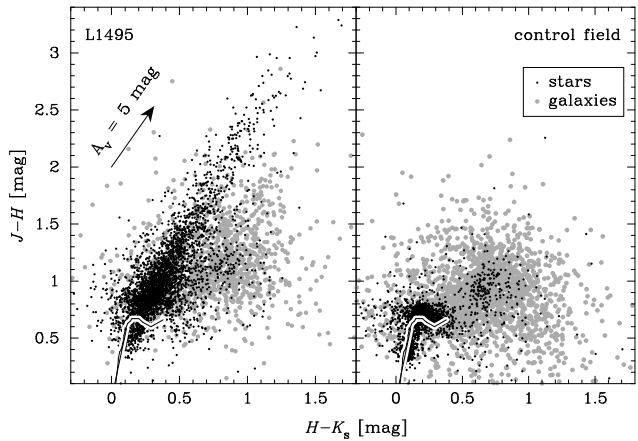}
	\caption{\label{fig04} NIR color-color diagram of the science ({\em left})
and the control field ({\em right}). A shift caused by an extinction of 
$A_\mathrm{V}=\unit[5]{mag}$ (assuming the reddening law of \citealp{rie85}) is
indicated by the arrow in the left panel.
The continuous line shows an unreddened Zero-Age Main-Sequence
\citep{sie00} for comparison.}
	\end{center}
\end{figure}

\begin{figure*}[tb]
	\includegraphics[height=7in,angle=-90]{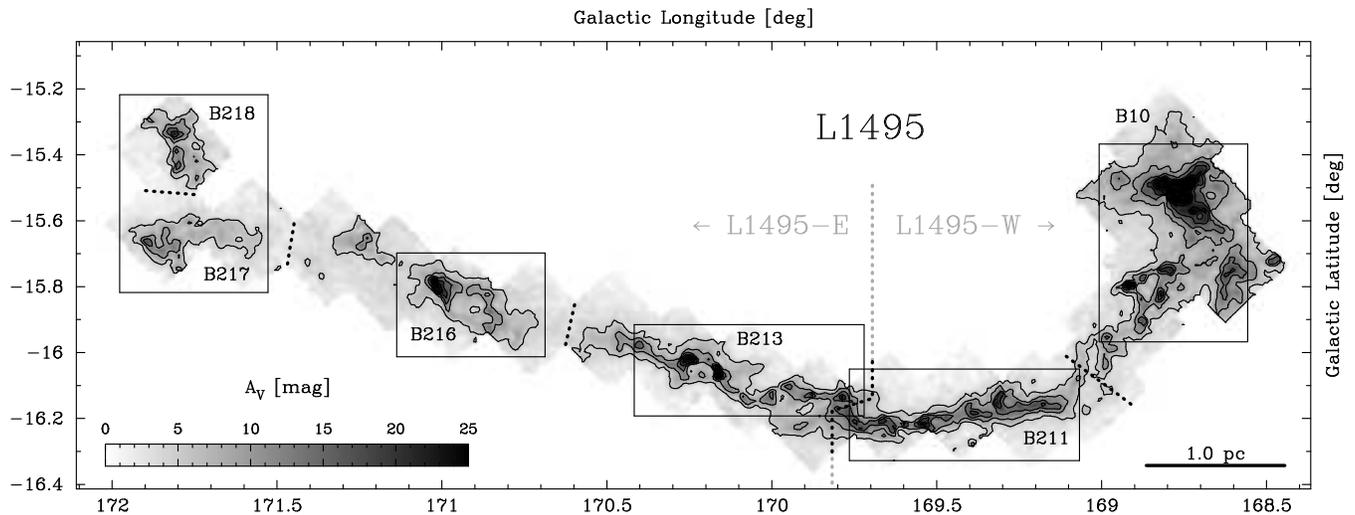}
	\caption{\label{fig05} Extinction map of the Filament
	L\,1495 with a resolution of $\unit[0.9]{\arcmin}$ derived from deep NIR
	observations with Omega2000. Contours are plotted in
	steps of $A_\mathrm{V}=\unit[5]{mag}$. We separate the filament into different
	subregions, which conform to Barnard's Dark Objects \citep{bar27a}.
	The boxes indicate the positions of
	the zoom-ins shown in Fig.\,\ref{fig06}.}
\end{figure*}

\begin{figure*}[bt]
	\includegraphics[height=7in,angle=-90]{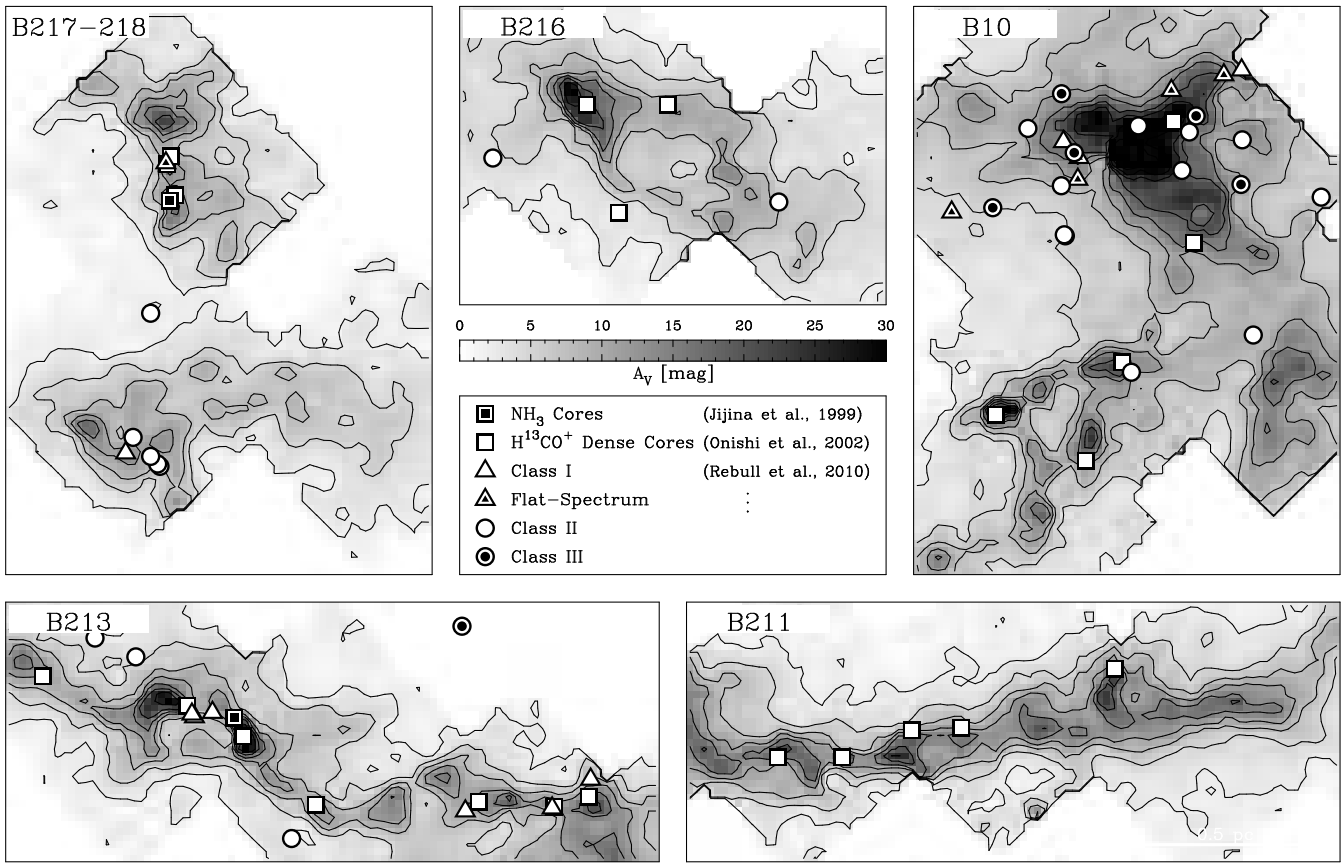}
	\caption{\label{fig06} Zoom-ins into the extinction map. The scale bar
	indicated in the bottom panel is common to all sub-plots. Contours are
	plotted in steps of 5$\sigma$ each. The positions of dense cores mapped in
	NH$_3$ \citep{jij99}, H$^{13}$CO$^+$ (\citetalias{oni02}) are overplotted
	alongside YSOs \citep{reb10}.}
\end{figure*}

We used the $JHK_\mathrm{s}$ photometry of the detected sources to derive
dust extinction through the Taurus filament L\,1495. In the following, we
first shortly describe the adopted technique and then present the results.

As mentioned earlier, in addition to stars there is a significant number of
background galaxies among detected sources. The intrinsic colors of galaxies
are very different from those of stars, and therefore theses two classes
need to be treated separately in deriving extinction towards them. This
difference is illustrated in Fig.\,\ref{fig04}, which shows color-color
diagrams of our reference (right panel) and one science field (left panel).
Especially in the reference field, one can clearly see the different color
distribution of stars and galaxies, as it
is unaffected by the dust extinction from the molecular cloud. We note that
there is a concentration of stars at $[H-K_\mathrm{s},J-H]=[0.7,1.0]$ that is
clearly outside the Zero-Age Main-Sequence (ZAMS) of stars. These sources
could be YSOs, brown dwarfs, or even point like distant galaxies and or quasars.

Recently, \citet{fos08} presented a NIR color-excess mapping technique
tailored specifically for inclusion of galaxies. In this technique, the
{\em colors of stars} behind the cloud are compared to the intrinsic colors
of stars in the reference field. In addition, the {\em magnitude
dependent colors} of {\em galaxies} are compared to those in the reference
field. This comparison yields a pencil-beam-like extinction value towards
each source \citep[for further details we refer to][]{fos08}. From these
values we then generate a uniformly gridded extinction map by smoothing
the data using a Gaussian kernel. Sigma-clipping removes sources with outliers
in extinction, which might be caused by intrinsically different colors
(e.g.\ YSOs, brown dwarfs, etc). For our map, we chose a resolution of
$\unit[0.9]{\arcmin}$ as a compromise between resolving the small
scale structures in the filament and having low noise over the dynamical
range covering most of the map. In this resolution, the map covered the
dynamical range of $A_\mathrm{V}=1.5\ldots\unit[35]{mag}$. The noise of the
final map depends on extinction, being $\sigma=\unit[0.5]{mag}$ at
$A_\mathrm{V}=\unit[0]{mag}$ and $\sigma=\unit[2]{mag}$ at
$A_\mathrm{V}=\unit[35]{mag}$.

Fig.\,\ref{fig05} shows the resulting extinction map, with blow-ups of 
different regions of it shown in Fig.\,\ref{fig06}. The map reveals, on one
hand, a remarkably well defined, high aspect ratio filamentary structure,
and on the other hand complex clumpy substructures within this filament.
We note that within the mapped region we find six Barnard objects,
namely the clumps B\,211, B\,213, B\,216, B\,217, and B\,218 (which represent
the {\em filament}), and B\,10 (which forms the {\em hub}).

Within the filament, the lowest completely mapped iso-contour is at
$A_\mathrm{V}=\unit[5]{mag}$. The total mass within this region is
$\unit[280\pm19]{\msun}$, yielding a mass surface density of
$\Sigma_\mathrm{M}=\unit[182\pm13]{\msun\,pc^{-2}}$.
Table\,\ref{tab01} lists these parameters also separately for the
subregions. The mass was estimated from the extinction map by assuming
all hydrogen in molecular form and the standard conversion factor of
$N_\mathrm{H_2}/A_\mathrm{V}=\unit[0.94\times10^{21}]{cm^{-2}mag^{-1}}$
\citep{boh78,rie85}, which then yields the total hydrogen mass
\begin{equation}
	\label{equ01}
	\frac{M_\mathrm{H}}{\msun} = 4.89\times10^{-3}\,\left( \frac{D}{\unit[137]{pc}}\right)^2\left(\frac{\theta}{\unit[0.45]{\arcmin}}\right)^2\,\sum_i \left(\frac{A_{{\rm V},i}}{\unit[1]{mag}}\right)
\end{equation}
where $D$ is the distance to the cloud, $\theta$ is the pixel scale of the
map, and $A_{{\rm V},i}$ are the extinction values of the individual pixels.
Using a standard cloud composition of 63\% hydrogen, 36\% helium and
1\% dust \citep{lom06} one can assume the total mass from Equation\,(\ref{equ01})
to be
\begin{equation}
	\label{equ02}
	M_\mathrm{tot}=1.37\,M_\mathrm{H}.
\end{equation}

\begin{deluxetable}{lccc}
\tablecaption{\label{tab01}Barnard Objects in the L\,1495 filament}
\tablehead{   
  \colhead{Region} &
  \colhead{Mass} &
  \colhead{Area} &
	\colhead{$\Sigma_\mathrm{M}$\tablenotemark{a}}\\
	\colhead{} &
  \colhead{[$\msun$]} &
	\colhead{[pc$^2$]} &
	\colhead{[$\msun$\,pc$^{-2}$]}
}
\startdata
B\,211 & $ 84\pm  5$ & $ 0.40\pm 0.06$ & $ 210.7\pm  14.8$  \\
B\,213 & $ 87\pm  6$ & $ 0.48\pm 0.07$ & $ 180.3\pm  12.6$  \\
B\,216 & $ 57\pm  4$ & $ 0.34\pm 0.05$ & $ 169.8\pm  11.9$  \\
B\,217 & $ 29\pm  2$ & $ 0.19\pm 0.03$ & $ 154.8\pm  10.8$  \\
B\,218 & $ 22\pm  1$ & $ 0.13\pm 0.02$ & $ 174.8\pm  12.2$  \\ \hline \\[-1ex]
Total & $280\pm 19$ & $ 1.54\pm 0.22$ & $ 182.3\pm  12.8$
\enddata
\tablenotetext{a}{Values for mass and area in the table are rounded, and
therefore their division leads to different values compared to the
correct calculation.}
\tablecomments{All parameters are calculated for regions with
\mbox{$A_\mathrm{V}\geq\unit[5]{mag}$}.}
\end{deluxetable}

The blow-ups shown in Fig.\,\ref{fig06} reveal numerous dense fragments
with peak extinction values of $A_\mathrm{V}\gtrsim\unit[15]{mag}$. This population
of {\em dense cores}\footnote{Our analysis (\S\ref{s-analysis_cores}) shows that
these are indeed dense cores, i.e. gravitationally bound entities with
densities of $n_{\mathrm{H}_2}\sim\unit[10^4]{cm^{-3}}$. Therefore,
we already introduce this term here, for reasons of consistency.}
will be investigated in more detail in the
following section. The main filament is strongly meandering along almost
its entire length of $\unit[8]{pc}$, and sometimes forms even
ring-like structures (e.g.\ B\,216, B\,10). However, B\,211 is different in
that it exhibits a linear, narrow {\em bar} with a length
of $\unit[1.5]{pc}$ and a median FWHM
of only $\unit[0.11]{pc}$. The inter-core regions also show considerably higher
extinction values of $A_\mathrm{V}\gtrsim\unit[10]{mag}$ in contrast to the
other parts of the filament, where cores seem to be rather isolated.

\subsection{The Dense Core Population}
\label{s-results_cores}

Identification and characterization of small-scale structures, such as cores,
from 2D data (e.g.\ extinction maps or continuum observations) is a
long-lasting, non-trivial problem, and several methods exist
\citep[e.g.,][]{stu90,clumpfind}.
We adopted a two-step approach, which consisted of
background removal and 2D thresholding.
In the first step, we performed a subtraction of large-scale structures from
the map using wavelet filtering. This resulted in a map describing structures
only at spatial scales smaller than $\unit[0.14]{pc}$ (``cores-only'' map,
c.f.\ \citealt{alv07}). In the second step,
we used the 2-dimensional thresholding algorithm
{\sc Clumpfind2D} \citep{clumpfind} to identify a population of dense cores
from the cores-only map. The algorithm works by first contouring the data
(usually these levels are defined by a detection threshold, and subsequent
contours with separations of multiples of the rms noise of the observations),
and then searches for peaks of emission, which mark the cores. It then follows
these down to lower intensities, assigning every pixel to a certain core
(if any). Obviously, such an algorithm works best for isolated cores.
\citet{pin09} pointed out that for observations, where one has to deal with
blending of cores, {\sc Clumpfind2D} is very sensitive to the
choice of input parameters.
Therefore, we ran a set of calculations with different values of detection
thresholds and level separations. We performed various tests, which included
visual verification of the positions, sizes and shapes of the cores. A detection
threshold and a level separation of $3\sigma$
was chosen to yield a representative core distribution. This is in agreement
with the simulations by \citet{kai09pipe}, in which such parameters resulted in
an acceptable core identification.

However, the narrow {\em bar} in B\,211 was not removed by the wavelet filter,
eventually not allowing a look at the ``naked cores''. This prohibited a
reliable recovery of the dense cores and their properties in B\,211. We believe
that this is caused by its different structure, which is discussed in more
detail in \S\,\ref{s-analysis_sf}. Therefore, we did not consider this region
in the following investigation. Furthermore, we also did not take into account
B\,10 because of its cluster-like nature and very high extinction, that does not
allow to break the structure into separate cores \citep{kai09pipe}.
Follow-up analysis using {\em dendrograms} \citep{ros08} on 3D
position-position-velocity molecular line data might allow us to also
disentangle the dense core population in these two regions in the western part
(L\,1495--W). In this paper, however, due to the limitations of 2D
data we restrict our analysis to the dense core population in the
eastern part (L\,1495--E), which consists of regions B\,213, B\,216,
B\,217, and B\,218.

\begin{figure*}[tb]
	\begin{center}
	\includegraphics[height=7in,angle=-90]{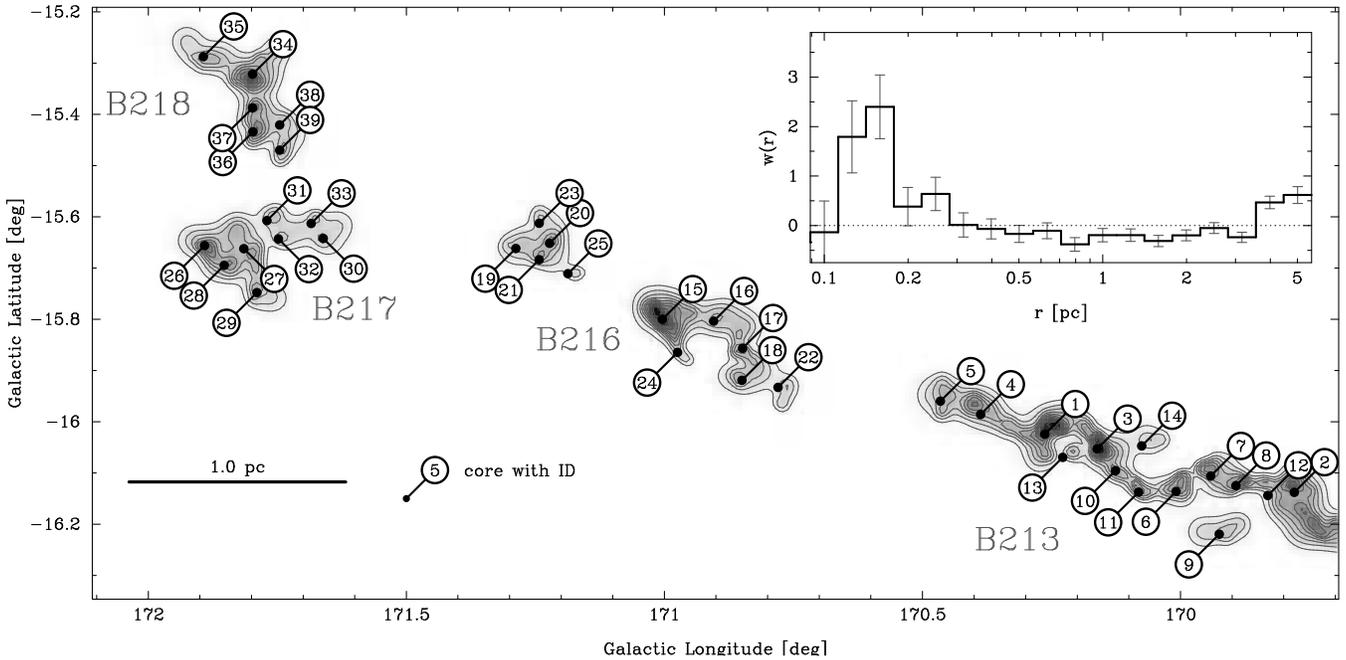}
	\caption{\label{fig07} The distribution of cores in L\,1495--E. The
	core-IDs can be found in Table\,\ref{tab02}. In the top right corner, the
	inset depicts the two-point correlation function $w(r)$.
	}
	\end{center}
\end{figure*}

In total 39~cores were detected in L\,1495--E
(Fig.\,\ref{fig07},
Table\,\ref{tab02}). We calculated the
core orientation together with the zeroth, first and second moment, which after
suitable conversions conformed to the core mass, position and FWHM along
the minor and major axis. Furthermore, we defined
the effective radius as $r_\mathrm{eff}=(A/\pi)^{1/2}$,
assuming spherical symmetry. We note that the mass is quite insensitive for the choice
of the threshold level, changing on average only 15\% when
going from 3$\sigma$ to 6$\sigma$. A few cores (\#23, \#29 and \#35) extended
beyond our mapped region. We used data from the 2MASS catalog to extend the
map to cover these cores completely.

The nearest neighbor separation of the cores exhibits a strong peak at
$\unit[0.13]{pc}$ with a dispersion of $\unit[0.04]{pc}$, and all cores
had at least one neighbor within $\unit[0.25]{pc}$. Another approach to
quantify the ``clustering'' is to measure the two-point
correlation function \citep[e.g.,][]{joh00,har02}, where
the number of core pairs $N$ with separations in the interval
$[\log(r),\log(r)+\dd{\log(r)}]$ is compared to the number $N_0$ in a
sample, where cores were randomly distributed over the whole region.
The two-point correlation function is then
\begin{align*}
	w(r)=\frac{N}{N_0}-1.
\end{align*}
Applying this to our sample of cores, strong correlations, i.e.\ $w(r)>0$,
on separations of $r\sim\unit[0.15]{pc}$ and $r\sim\unit[5]{pc}$ are evident
(Fig.\,\ref{fig07}). The latter, large separation derives from small
clusters of dense cores at both ends of L\,1495--E, whereas the first one
represents the separation between the individual cores inside these clusters.
At our pixel scale of the map, a separation of $\unit[0.15]{pc}$ corresponds to
$\sim\unit[7]{px}$, and is, therefore, well above the resolution limit.

\begin{figure}[tb]
	\begin{center}
	\includegraphics[height=3.3in,angle=-90]{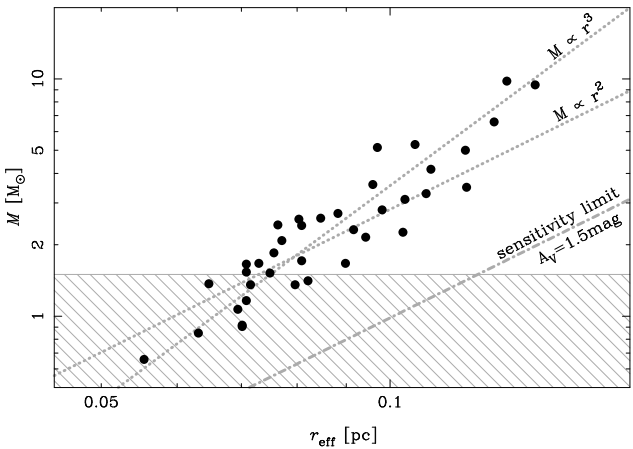}
	\caption{\label{fig08} Mass-size relation for our cores. The hatched area
	marks the region of incompleteness. The dotted lines represent the cases
	of constant volume density $M\propto r^3$ and column density $M\propto r^2$.}
	\end{center}
\end{figure}

\begin{figure}[tb]
	\begin{center}
	\includegraphics[height=3.3in,angle=-90]{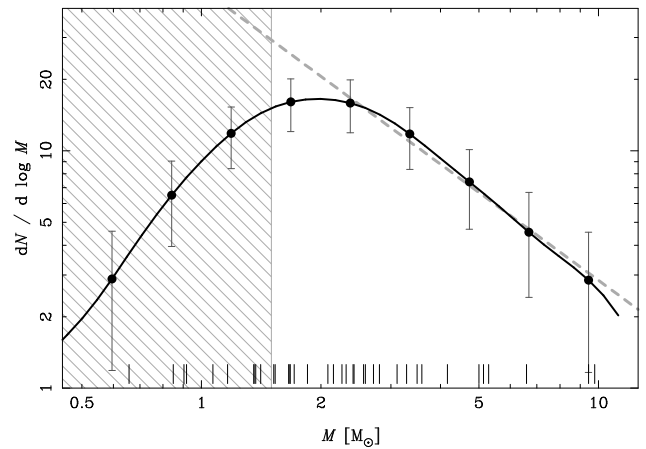}
	\caption{\label{fig09}The Dense Core Mass Function (DCMF) of L\,1495--E,
	which was obtained by smoothing the core masses
	(indicated as vertical dashes along
	the abscissa) with a Gaussian kernel with FWHM=$\unit[0.3]{dex}$. Points with
	error bars are plotted in separations of 50\% of the FWHM.
	The dashed line shows the power law
	fit with slope $\Gamma=1.2\pm0.2$ for $M>\unit[2.0]{\msun}$.
	The hatched area indicates the region of incompleteness.}
	\end{center}
\end{figure}

Core masses range from $M_\mathrm{core}=0.4\ldots\unit[10]{\msun}$.
The total mass found in cores is $\unit[104\pm16]{\msun}$, which makes
up $\sim$50\% of the total mass in L\,1495--E (as defined by the
$A_\mathrm{V}=\unit[5]{mag}$ contour). We derived mean core densities
$\rho=3\,M_\mathrm{core}/(4\,r_\mathrm{eff}^3)$, and hydrogen number densities
$n_{\mathrm{H}_2}=\rho/(\mu\,m_\mathrm{H})$, where $m_\mathrm{H}$ is
the proton mass and $\mu=2.34$ the mean molecular weight per hydrogen molecule,
assuming a molecular cloud with standard composition \citep{lad08}.
The frequency distribution of the derived number densities closely follows
a Gaussian distribution with
$n_{\mathrm{H}_2}=\unit[14.3\times 10^3]{cm^{-3}}$ and a dispersion of
$\unit[4.1\times10^3]{cm^{-3}}$.

In Fig.~\ref{fig08}, we show the mass-size relation of the cores in our sample.
For reference, the figure also shows the sensitivity limit of
$A_\mathrm{V}=\unit[1.5]{mag}$ for an isolated, flat and spherical core.
\citet{lad08} derived a slope of of $M\propto r_\mathrm{eff}^{2.6}$ for the
dense cores of the Pipe Nebula. Due to the low number of cores and the strong
influence of the completeness limit on the fit, we rather show
representative relations with constant column density
($M\propto r_\mathrm{eff}^2$), which would be
predicted for clouds obeying Larson's laws \citep{lar81},
and constant volume density ($M\propto r_\mathrm{eff}^3$), which
would be expected in case of constant thermal pressure and kinetic temperature.
However, with the data at hand we cannot rule out one or the other model.

The Dense Core Mass Function (DCMF), which is the number of cores per
logarithmic mass interval $\dd{N}/\dd{\log(m)}\propto m^{-\Gamma}$, 
exhibits its maximum at $\sim\unit[2.0]{\msun}$, and
falls off to both the low-mass and high-mass ends (Fig.\,\ref{fig09}).
This peak could be either true, or simply be caused due to incompleteness.
However, estimating the completeness and accuracy of the DCMF
is not a trivial task
\citep[see e.g.,][]{kai09pipe,pin09}. This is partly due to the inability of the
adopted algorithm to detect cores from the variable, more extended column
density component. Partly it is also due to the noise in the map, which makes
mass determination of low-mass cores significantly less accurate than for
high-mass cores. In an analysis similar to this paper, \citet{kai09pipe}
analyzed the accuracy derived for the cores in the Pipe Nebula, and concluded
the DCMF to be accurate above about $\unit[1.2]{\msun}$. By comparing the noise
levels within these two maps we roughly estimate the DCMF to be accurate for
$M_\mathrm{core}\geq\unit[1.5]{\msun}$. A power-law fit to the DCMF
above this completeness
limit yields an exponent $\Gamma=1.2\pm0.2$. This slope agrees with the results
of \citetalias{oni02}, which found $\Gamma=1.5\pm0.3$ for
$3<M/\msun<20$. Slopes of $\Gamma\sim1.2$ are also commonly observed in
other star-forming regions \citep[e.g.][]{mot98,rat09,and10}.

\begin{figure}[tb]
	\begin{center}
	\includegraphics[height=3.3in,angle=-90]{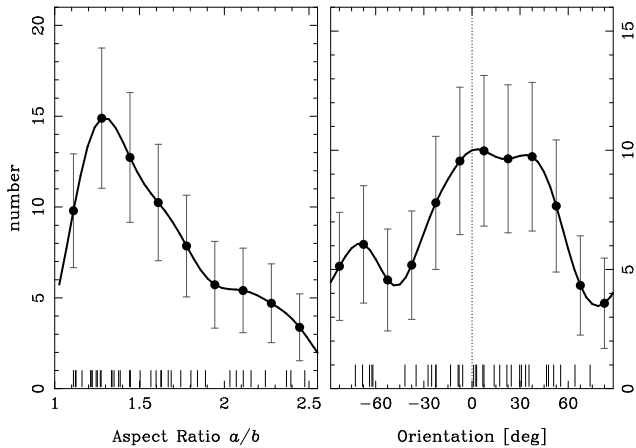}
	\caption{\label{fig10}{\em left panel:} Distribution of aspect ratios $R$,
	which was obtained by smoothing the data points (indicated as vertical
	dashes on the abscissa) with a Gaussian kernel of FWHM$=0.3$. Data points are
	plotted in steps of 50\% of the FWHM. ---
	{\em right panel:} Distribution of core orientations
	with respect to the filament. The FWHM for the Gaussian kernel was
	chosen to be $\unit[30]{^\circ}$.}
	\end{center}
\end{figure}

The distribution of aspect ratios $R$, which is defined as the
ratio of the major vs.\ the minor axis $a/b$, is strongly skewed. The mode
of the distribution is found to be $R=1.3$, whereas the mean aspect ratio is
found to be $R=1.5$, with a dispersion of $0.4$ (Fig.\,\ref{fig10}, {\em left panel}).
Within the region of our observations, we find five cores
from the sample of optically selected cores from \citet{lee99}.
Three of these cores were classified as round
($R=1$), and the remaining two had only moderate aspect ratios of $R\sim1.8$.
Compared to the all-sky sample of 406~cores from \citet{lee99} with
$R=2.4\pm0.1$, the cores in Taurus are therefore only moderately
elongated.

We also considered the orientation of the cores with respect to the filament.
Due to its meandering nature, this is not a trivial task. Therefore,
we first defined the baseline of the filament by setting points
along its path, and then interpolated between these points using a cubic spline
function. The orientation angle of the core with respect
to this baseline was then obtained, and we find that the 39 cores seem to
have a slight tendency to align with the filament. However, a KS-test results
in a probability of $p=0.46$ that this distribution is drawn from a uniform
distribution. Given the large error bars on the individual orientations, 
which come along with the small aspect ratios, we can therefore not
draw any secure conclusions about preferential or random core alignment 
in this filament.

\begin{deluxetable}{lcccccc}
\tablecolumns{7}
\tablecaption{\label{tab02}Dense Cores in L\,1495--E}
\tablehead{
	\colhead{ID} &
	\colhead{$l$} &
	\colhead{$b$} &
	\colhead{$M_\mathrm{core}$} &
	\colhead{$r_\mathrm{eff}$} &
	\colhead{$n_{\mathrm{H}_2}$} &
	\colhead{Aspect}\\
	\colhead{\#} &
	\colhead{[deg]} &
	\colhead{[deg]} &
	\colhead{[$\msun$]} &
	\colhead{[pc]} &
	\colhead{[$\unit[10^4]{cm^{-3}}$]} &
	\colhead{Ratio $R$}
}
\startdata
\cutinhead{B\,213}
 1 & 170.26 & -16.02 &   9.4 & 0.14 &  1.4 &  1.6 \\
 2 & 169.78 & -16.14 &   5.3 & 0.11 &  1.8 &  1.2 \\
 3 & 170.16 & -16.05 &   5.1 & 0.10 &  2.3 &  1.5 \\
 4 & 170.39 & -15.99 &   5.0 & 0.12 &  1.2 &  1.7 \\
 5 & 170.47 & -15.96 &   3.1 & 0.10 &  1.2 &  1.1 \\
 6 & 170.01 & -16.14 &   2.7 & 0.09 &  1.6 &  1.2 \\
 7 & 169.94 & -16.11 &   2.6 & 0.08 &  1.8 &  1.1 \\
 8 & 169.89 & -16.12 &   2.4 & 0.08 &  1.9 &  1.2 \\
 9 & 169.93 & -16.22 &   2.3 & 0.10 &  0.9 &  2.1 \\
 10 & 170.13 & -16.10 &   1.8 & 0.08 &  1.8 &  1.3 \\
 11 & 170.08 & -16.14 &   1.7 & 0.07 &  1.8 &  1.7 \\
 12 & 169.83 & -16.14 &   1.5 & 0.07 &  1.8 &  2.1 \\
 13 & 170.23 & -16.07 &   1.4 & 0.08 &  1.1 &  2.6 \\
 14 & 170.08 & -16.05 &   0.9 & 0.07 &  1.1 &  1.2 \\
\cutinhead{B\,216}
 15 & 171.00 & -15.80 &   9.8 & 0.13 &  1.7 &  1.3 \\
 16 & 170.90 & -15.80 &   3.3 & 0.11 &  1.0 &  1.7 \\
 17 & 170.85 & -15.86 &   2.8 & 0.10 &  1.2 &  1.6 \\
 18 & 170.85 & -15.92 &   2.3 & 0.09 &  1.2 &  1.6 \\
 19 & 171.29 & -15.66 &   1.7 & 0.08 &  1.3 &  1.2 \\
 20 & 171.22 & -15.65 &   1.7 & 0.07 &  1.9 &  1.2 \\
 21 & 171.24 & -15.68 &   1.4 & 0.06 &  2.1 &  1.3 \\
 22 & 170.78 & -15.93 &   0.9 & 0.07 &  1.1 &  1.6 \\
 23 & 171.24 & -15.61 &   0.8 & 0.06 &  1.4 &  1.3 \\
 24 & 170.97 & -15.86 &   0.7 & 0.06 &  1.6 &  1.4 \\
 25 & 171.19 & -15.71 &   0.4 & 0.05 &  1.5 &  2.1 \\
\cutinhead{B\,217}
 26 & 171.89 & -15.66 &   4.2 & 0.11 &  1.3 &  1.2 \\
 27 & 171.81 & -15.66 &   3.6 & 0.10 &  1.7 &  2.4 \\
 28 & 171.85 & -15.69 &   2.6 & 0.08 &  2.0 &  1.3 \\
 29 & 171.79 & -15.75 &   2.2 & 0.09 &  1.1 &  1.5 \\
 30 & 171.66 & -15.64 &   1.7 & 0.09 &  1.0 &  1.8 \\
 31 & 171.77 & -15.61 &   1.4 & 0.08 &  1.1 &  2.0 \\
 32 & 171.75 & -15.64 &   1.2 & 0.07 &  1.4 &  1.7 \\
 33 & 171.68 & -15.61 &   1.1 & 0.07 &  1.3 &  2.5 \\
\cutinhead{B\,218}
 34 & 171.80 & -15.32 &   6.6 & 0.13 &  1.3 &  1.6 \\
 35 & 171.89 & -15.29 &   3.5 & 0.12 &  0.8 &  2.1 \\
 36 & 171.80 & -15.43 &   2.4 & 0.08 &  2.2 &  1.1 \\
 37 & 171.80 & -15.39 &   2.1 & 0.08 &  1.9 &  1.3 \\
 38 & 171.75 & -15.42 &   1.5 & 0.07 &  1.5 &  1.4 \\
 39 & 171.75 & -15.47 &   1.4 & 0.07 &  1.5 &  1.3
\enddata
\end{deluxetable}

\subsection{Reddening Law}
\label{s-reddening}

Algorithms like {\sc gnicer} obtain the excess extinction of
a source by comparing its position in the color-color diagram with respect
to a control field. It is crucial
for a proper estimation of extinction values to assume a reddening law
\citep[e.g.,][]{rie85,ind05}. For observations in $JHK_\mathrm{s}$
the ratio of the colour excesses
\begin{equation}
	\alpha=\frac{E(J-H)}{E(H-K_\mathrm{s})}
\end{equation}
is constant, and defines the direction of the displacement of a source
in the color-color diagram when it is subject to extinction.
Therefore, the colors $(J-H)$ and $(H-K_\mathrm{s})$ are connected
by the linear relation
\begin{equation}
	\label{equ03}
 (J-H)=\alpha~(H-K_\mathrm{s})+\beta.
\end{equation}
As unreddened ZAMS stars occupy only a small region of
the color-color-diagram, the slope $\alpha$ could be simply
obtained by fitting Equation\,(\ref{equ03}) for all stars simultaneously. This,
however, would give the high number of low-extinction stars too much weight
compared to the low number of high-extinction stars. Therefore, we followed a
procedure similar to \citet{lom06}.

The first step
was to obtain visual extinctions $A_\mathrm{V}$ of every star
by converting color excesses $E(J-H)$ and $E(H-K_\mathrm{s})$ assuming the
reddening law of \citet{ind05}. Stars are then binned in the
color-color diagram according to their extinction $A_\mathrm{V}$. These data points
are then fitted according to Equation\,(\ref{equ03}) to obtain a
corrected slope. With the aid of this new slope, corrected extinction
values $A_\mathrm{V}$ are then computed. This procedure is repeated
until the slope $\alpha$ converges, which generally happened after 3-6
iterations. In order to decrease dependence on the bin size, we repeated this
procedure for bin sizes ranging from
$\Delta A_\mathrm{V}=\unit[0.5]{mag}\ldots\unit[1.5]{mag}$
in steps of $\unit[0.2]{mag}$. The best fit slope $\alpha$ was then the
variance weighted mean from all runs with different bin sizes.

For this analysis we considered only sources classified as stars without any
photometry flags. The best fit slope was found to be
$\alpha=1.82\pm0.08$. This value is in very good agreement to the reddening
laws of \citet[$\alpha=1.82\pm\unit0.03$]{lom06} and
\citet[$\alpha=1.78\pm0.15$]{ind05}.


\section{Discussion}
\label{s-discussion}

\subsection{Star Formation in the Filament}
\label{s-analysis_sf}

Due to its proximity, the Taurus Molecular Cloud is an
excellent target to study the connection between filamentary structure
and star formation.
In this paper we present a new, high-resolution view of the L\,1495 filament,
allowing characterization of its density structure down to tenth-of-a-parsec
scales. Indeed, as illustrated in Figs.\,\ref{fig05}-\ref{fig07},
L\,1495 shows a highly fragmented small-scale structure and in some
places extinctions, which are higher than can be traced by the chosen
technique (i.e.\ $A_\mathrm{V}\gtrsim\unit[35]{mag}$).

To estimate the star-forming potential of the filament, we
calculated the mass-per-length $M_\mathrm{line}$. For an undisturbed
filament, filaments with a mass exceeding the critical value of
$M_\mathrm{line}^\mathrm{crit}=\unit[15]{\msun\,pc^{-1}}$ are prone to
filamentary fragmentation and collapse \citep{ost64,inu97}. In presence
of magnetic fields, this critical value can be increased by a
factor of 2 \citep{fie00a}. Due to the presence of dense cores,
L\,1495 does not really fulfill the requirement of being an undisturbed filament.
However, considering only the large scale structure, which was filtered out
in the wavelet transform, we derived a value of
$M_\mathrm{line}=\unit[16.8\pm0.2]{\msun\,pc^{-1}}$
for the whole filament. As this must be seen as a lower limit, as some of the
mass is already bound in the dense cores, this value underlines the
star-forming potential of the filament.

In Fig.\,\ref{fig06} we show the previously known
YSOs\footnote{\citet{reb10} separate
YSOs into Class~I, Flat-Spectrum, Class~II and Class~III. We include all
YSOs, which were classified as {\em confirmed}, {\em probable}, and
{\em possible} Taurus members in their paper.}. Interestingly, while YSOs
are generally found close to dense cores in the filament, the bar-like
B\,211 is completely devoid of them \citep{gol08}.
Given that the mass-per-length value in this particular region is also high
($M_\mathrm{line}=\unit[15]{\msun\,pc^{-1}}$), and that it also
harbors several H$^{13}$CO$^+$ cores signifying the presence
of high density gas ($n_\mathrm{crit}=\unit[10^5]{cm^{-3}}$),
it can be argued to be well underway towards star formation.

\begin{figure}[tb]
	\begin{center}
	\includegraphics[height=3.3in,angle=-90]{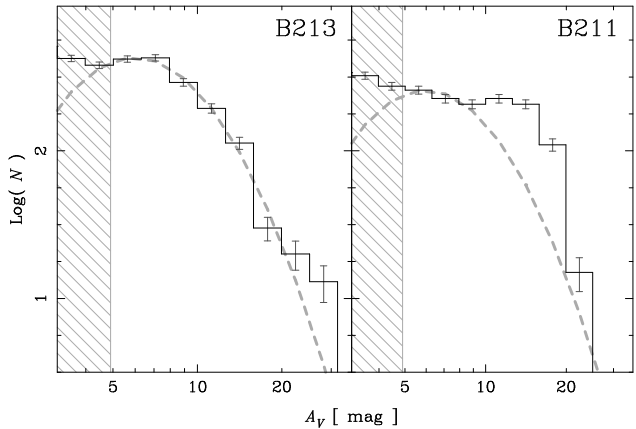}
	\caption{\label{fig11} Probability Density Function
	(PDF) of B\,213 and B\,211. Our observations do not completely cover regions
	with $A_\mathrm{V}<\unit[5]{mag}$ (hatched area). The dashed line
	is the log-normal fit to B\,213, which is shown as a scaled-down version
	in the right panel.}
	\end{center}
\end{figure}

\begin{figure}[tb]
	\begin{center}
	\includegraphics[height=3.3in,angle=-90]{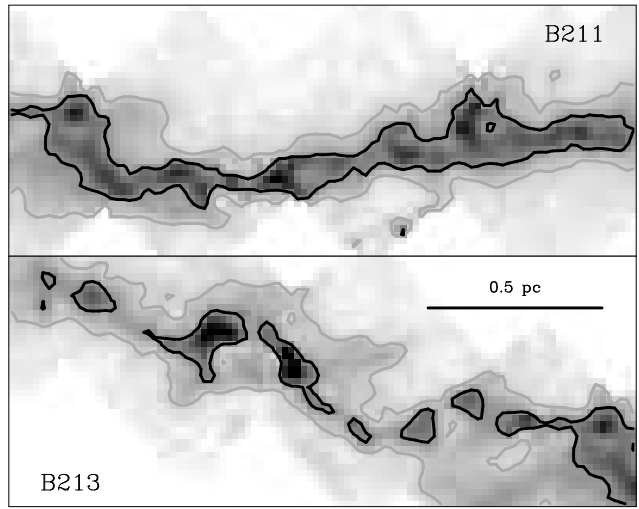}
	\caption{\label{fig12} Zoom-ins into the regions B\,211 and B\,213 with
	contours at $A_\mathrm{V}=\unit[5]{mag}$ (gray) and
	$A_\mathrm{V}=\unit[11]{mag}$ (black).}
	\end{center}
\end{figure}

More insight into the assembly of material in the non-star-forming part of the
filament can be gained by examining the distribution of column densities.
Therefore, in Fig.\,\ref{fig11} we compare the quiescent B\,211 with B\,213,
in which star formation has already occurred. The column density distribution of
B\,213 shows a continuous decline towards higher extinctions.
In contrast, B\,211 shows a flat plateau
at intermediate extinctions, falling steeply at extinctions higher than
$A_\mathrm{V}\sim\unit[13]{mag}$. This excess with respect to the column density
distribution of B\,213 can be related particularly to {\em inter-core} regions.
This is illustrated in Fig.\,\ref{fig12}, which shows
the contour of $A_\mathrm{V}=\unit[11]{mag}$ for both B\,211 and B\,213.
Clearly, in B\,213 such extinction range is related to cores, while
in B\,211 it is more continuous and related to the material {\em surrounding}
the cores. Therefore, we suggest that B\,211 is not at all devoid of
star-forming capability, but in an early stage of its evolution where the
filament is fragmenting and the inter-core regions have not yet been cleared
from material by accretion towards the cores. Given the high mass reservoir
in the filament, such fragmentation is inevitably leading to star formation
in B\,211.

In L\,1495--E, the star-forming part of the filament, we investigated
the spatial distribution of YSOs with respect to the filament.
Class~I and Flat-Spectrum sources are found at extinctions of
$A_\mathrm{V}=\unit[8.2\pm3.0]{mag}$, which is higher compared to the
extinctions at the positions of
Class~II and III ($A_\mathrm{V}=\unit[5.2\pm2.6]{mag}$). The mean spatial
separation of Class~I and Flat-Spectrum sources from the filament was found
to be only $\unit[0.06]{pc}$, whereas older sources showed considerably larger
mean distances. To verify whether this could be a pure evolutionary
effect, we performed a simple Monte Carlo simulation to model the time
dependence of the spatial distribution of sources. We distributed 10,000 test
sources with a 3D velocity dispersion of
$\unit[0.2]{km\,s^{-1}}\sim\unit[0.2]{pc\,Myr^{-1}}$
along the baseline of the filament
(defined in \S\ref{s-results_cores}). The mean distance of sources from the
filament was then measured as a function of time.
The test sources started to show larger mean distances than Class~I and
Flat-Spectrum sources after $\sim\unit[1]{Myr}$. A mean distance similar to
that of Class~II and Class~III objects was reached after
$\sim\unit[3.5]{Myr}$, which is in good agreement with the evolutionary
timescale of these sources \citep{luh03,luh09}.

\subsection{The Nature of the Dense Cores}
\label{s-analysis_cores}

\begin{figure}[tb]
	\begin{center}
	\includegraphics[height=3.3in,angle=-90]{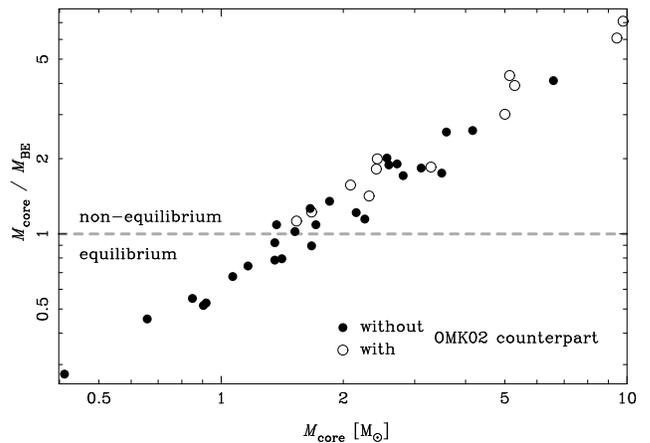}
	\caption{\label{fig13} Ratio of core masses $M_\mathrm{core}$ to the
	Bonnor-Ebert critical mass $M_\mathrm{BE}$ plotted against core mass.
	$M_\mathrm{BE}$ was calculated individually for every core, assuming its
	respective density and a temperature of $T=\unit[10]{K}$. Open circles
	represent cores which have a counterpart from the dense core sample of
	\citetalias{oni02}.}
	\end{center}
\end{figure}

In \S\ref{s-results_cores}, we defined and examined the dense core population
in L\,1495--E. To test whether these cores are gravitationally
bound entities, one can estimate the Bonnor-Ebert critical mass \citep{lad08}
\begin{align*}
	M_\mathrm{BE} &\sim 1.82\,\left(\frac{n_{\mathrm{H}_2}}{\unit[10^4]{cm^{-3}}}\right)^{-1/2}\,\left(\frac{T}{\unit[10]{K}}\right)^{3/2}\,\msun.
\end{align*}
Within this framework, in cores with $M_\mathrm{core}>M_\mathrm{BE}$
thermal motions do not provide enough support against gravity, and in the
absence of other forces they inevitably collapse and form protostars.
With a
mean density of $n_{\mathrm{H}_2}=\unit[1.5\times10^4]{cm^{-3}}$ and a
temperature of $T=\unit[10]{K}$ \citep{ter10}, we obtain
$M_\mathrm{BE}=\unit[1.5]{\msun}$. Calculating the Bonnor-Ebert critical mass
for all cores allowed us to analyze the stability of each
core separately (Fig.\,\ref{fig13}). As a result we find that in
L\,1495--E, the majority of cores is prone to collapse. Without exception,
all cores that coincide with a dense H$^{13}$CO$^+$ core \citepalias{oni02}
exhibit $M_\mathrm{core}>M_\mathrm{BE}$. Apparently, in these cores the collapse
has already led to a density enhancement, which can be traced by
H$^{13}$CO$^+$ with its critical density of $n_\mathrm{crit}\sim\unit[10^5]{cm^{-3}}$.

Compared e.g.\ to the starless core population in the Pipe Nebula
($n_{\mathrm{H}_2}=\unit[7.3\times10^3]{cm^{-3}}$, \citealt{lad08}), the cores
in the Taurus filament are clearly denser by a factor of 2. For the cores in the
Pipe Nebula, the Bonnor-Ebert critical mass is $\sim\unit[2]{\msun}$
(assuming $T=\unit[10]{K}$), and only a few cores exceed this value and
are prone to collapse. Cores in Taurus are also considerably more
clustered than those in the Pipe Nebula. The nearest neighbor
separation and two-point correlation function (Fig.\,\ref{fig07})
show a clear peak at separations of the order of $\unit[0.13]{pc}$, whereas cores in
the Pipe Nebula show separations of $\unit[0.38]{pc}$ \citep{rat09}.
Fragmentation
in Taurus happens mainly only on a very distinct spatial scale.
In the case of thermal fragmentation, such a preferential length scale
would be determined by the Jeans length
\begin{equation}
	\label{equ04}
	\lambda_J=c_s\,\left(\frac{\pi}{G\,\rho}\right)^{1/2},
\end{equation}
where $c_s$ is the local sound speed, $G$ the gravitational constant, and
$\rho$ the mass density. Assuming $c_s=\unit[0.19]{km\,s^{-1}}$
($T=\unit[10]{K}$) and the mean mass density of our
sample of cores ($\rho=\unit[5.8\times10^{-26}]{g\,cm^{-3}}$)
we derive $\lambda_J=\unit[0.18]{pc}$, which is only
slightly larger than the peak in the nearest neighbor distribution
of $\unit[0.13]{pc}$. This is in contrast to the Pipe Nebula with a separation
of $\unit[0.38]{pc}$ \citep{rat09}, which exceeds its local Jeans length
($\lambda_J\sim\unit[0.25]{pc}$).

We note that due to the unknown inclination angle $i$ we can only
derive the {\em projected} nearest neighbor separation, but the true
separation between cores remains unclear. For $i\sim45^\circ$ the true nearest
neighbor separation would match the local Jeans length. VLBA observations
allow accurate distance measurements of selected stars in the Taurus
Molecular Cloud exist \citep{loi07, tor07, tor09}, but do not allow to
estimate of the true spatial orientation of the filament.

We note, that non-zero inclination angles would not only affect the
derived nearest neighbor separation, but also the aspect ratio. Randomly
oriented cores would yield the observed mean aspect ratio of $R=1.5$,
if the true aspect ratio is $R_\mathrm{true}\sim1.7$ in the case of prolate,
and $R_\mathrm{true}\sim2.2$ for oblate cores, respectively \citep{mye91}.
However, from a large set of dense cores and Bok globules, \citet{ryd96} argues
that cores are generally rather prolate than oblate. This shape is furthermore
supported by theoretical simulations \citep{fie00b}, where helical magnetic
fields are able to produce a population of prolate dense cores through
filamentary fragmentation. The aspect ratios of these cores are in good
agreement with the sample of \citet{ryd96}, who finds $R=2.0\ldots2.5$.
Therefore, under the assumption of prolate cores, we find cores, which are
slightly more spherical ($R_\mathrm{true}\sim 1.7$). The same trend
is visible, when comparing the Taurus cores from the sample of
\citet{lee99} to their galactic average. Whether this is truly the case,
or simply a projectional effect cannot be determined. Due to the core
formation in the filament, we do not expect our sample of cores to be
oriented randomly, but rather inheriting some preferential direction.
However, the meandering nature of the filament does not allow
to determine a common inclination angle. Considering these limitations,
the elongation of our cores agrees reasonably well with
theoretical predictions and other observations.


\section{Summary}
\label{s-summary}

In this paper, we study the star formation in the Taurus filament L\,1495. In
particular we present a NIR extinction map of this filament with
unprecedented dynamical range ($A_\mathrm{V}\sim1.5\ldots\unit[35]{mag}$) and
resolution ($\unit[0.9]{\arcmin}$). The main
results of this paper are as follows:
\vspace{-1ex}
\begin{enumerate}
	\setlength{\itemsep}{-1ex}
	\item The extinction map of L\,1495 reveals the highly fragmented nature
	of the filament, harboring a population of dense cores preferentially
	separated by the local Jeans length. The mass-per-length in the filament is
	$M_\mathrm{line}=\unit[17]{\msun\,pc^{-1}}$, indicative of its star-forming
	potential. We find, that the part of the filament that harbors no YSOs,
	namely B\,211, shows an internal structure different from that of the rest
	of the filament. We argue, that B\,211 is still younger than other parts of
	the filament, being still in process of filament fragmentation. Given its
	high mass reservoir, star formation will inevitably ensue.
	\item The dense core population in L\,1495--E was
	investigated in detail. A total of 39 dense cores with masses between
	$M_\mathrm{core}=0.4\ldots\unit[10]{\msun}$ and densities of
	$n_{\mathrm{H}_2}=\unit[1.5\times10^4]{cm^{-3}}$
	were found. The majority of these cores exceeds the critical
	mass for collapse $M_\mathrm{BE}$, and is therefore prone to collapse.
	The high mass tail of the DCMF can be fitted with a power-law with an
	exponent $\Gamma=1.2\pm0.2$, a form commonly observed also in other
	star-forming regions.
\end{enumerate}


\acknowledgments

We thank Calar Alto Observatory for allocation of director's discretionary time
to this programme. This publication makes use
of data products from the Two Micron All Sky Survey \citep{2mass},
which is a joint project of the University of Massachusetts and the Infrared
Processing and Analysis Center/California Institute of Technology, funded by the
National Aeronautics and Space Administration and the National Science
Foundation. We furthermore acknowledge the use of NASA's SkyView facility
(\verb+http://skyview.gsfc.nasa.gov+)
located at NASA Goddard Space Flight Center \citep{mcg98} and the ViZier
database located at CDS in Strasbourg, France \citep{vizier}.


{\it Facilities:} \facility{CAO:3.5m (Omega2000)}


\begin{thebibliography}{}
\bibitem[Alves et al.(2007)]{alv07} Alves, J., Lombardi, M., \& Lada, C.~J.\ 2007, \aap, 462, L17
\bibitem[Andr{\'e} et al.(2010)]{and10} Andr{\'e}, P., et al.\ 2010, \aap, 518, L102 
\bibitem[Barnard et al.(1927)]{bar27a} Barnard, E.~E., Frost, E.~B., \& Calvert, M.~R.\ 1927, Carnegie institution of Washington, 1927.,
\bibitem[Barnard(1927)]{bar27b} Barnard, E.~E.\ 1927, Chicago: University of Chicago Press, 1927,  
\bibitem[Bergin \& Tafalla(2007)]{ber07} Bergin, E.~A., \& Tafalla, M.\ 2007, \araa, 45, 339 
\bibitem[Bertin \& Arnouts(1996)]{sextractor} Bertin, E., \& Arnouts, S.\ 1996, \aaps, 117, 393
\bibitem[Bohlin et al.(1978)]{boh78} Bohlin, R.~C., Savage, B.~D., \& Drake, J.~F.\ 1978, \apj, 224, 132 
\bibitem[Cantiello et al.(2005)]{can05} Cantiello, M., Blakeslee, J.~P., Raimondo, G., Mei, S., Brocato, E., \& Capaccioli, M.\ 2005, \apj, 634, 239 
\bibitem[Dobashi et al.(2005)]{dob05} Dobashi, K., Uehara, H., Kandori, R., Sakurai, T., Kaiden, M., Umemoto, T., \& Sato, F.\ 2005, \pasj, 57, 1
\bibitem[Fassbender(2003)]{fas03} Fassbender, R., PhD thesis, 2003, Universit\"at Heidelberg
\bibitem[Fiege \& Pudritz(2000a)]{fie00a} Fiege, J.~D., \& Pudritz, R.~E.\ 2000, \mnras, 311, 85 
\bibitem[Fiege \& Pudritz(2000b)]{fie00b} Fiege, J.~D., \& Pudritz, R.~E.\ 2000, \apj, 534, 291
\bibitem[Foster et al.(2008)]{fos08} Foster, J.~B., Rom{\'a}n-Z{\'u}{\~n}iga, C.~G., Goodman, A.~A., Lada, E.~A., \& Alves, J.\ 2008, \apj, 674, 831
\bibitem[Goldsmith et al.(2008)]{gol08} Goldsmith, P.~F., Heyer, M., Narayanan, G., Snell, R., Li, D., \& Brunt, C.\ 2008, \apj, 680, 428 
\bibitem[Gooch(1996)]{koords} Gooch, R.\ 1996, Astronomical Data Analysis Software and Systems V, 101, 80
\bibitem[Goodman et al.(2009)]{goo09} Goodman, A.~A., Pineda, J.~E., \& Schnee, S.~L.\ 2009, \apj, 692, 91 
\bibitem[Hartmann(2002)]{har02} Hartmann, L.\ 2002, \apj, 578, 914 
\bibitem[Henning et al.(2010)]{hen10} Henning, T., Linz, H., Krause, O., Ragan, S., Beuther, H., Launhardt, R., Nielbock, M., \& Vasyunina, T.\ 2010, \aap,518, L95
\bibitem[Indebetouw et al.(2005)]{ind05} Indebetouw, R., et al.\ 2005, \apj, 619, 931 
\bibitem[Inutsuka \& Miyama(1997)]{inu97} Inutsuka, S.-I., \& Miyama, S.~M.\ 1997, \apj, 480, 681
\bibitem[Jijina et al.(1999)]{jij99} Jijina, J., Myers, P.~C., \& Adams, F.~C.\ 1999, \apjs, 125, 161 
\bibitem[Johnstone et al.(2000)]{joh00} Johnstone, D., Wilson, C.~D., Moriarty-Schieven, G., Joncas, G., Smith, G., Gregersen, E., \& Fich, M.\ 2000, \apj, 545, 327
\bibitem[Kainulainen et al.(2006)]{kai06} Kainulainen, J., Lehtinen, K., \& Harju, J.\ 2006, \aap, 447, 597 
\bibitem[Kainulainen et al.(2007)]{kai07} Kainulainen, J., Lehtinen, K., V{\"a}is{\"a}nen, P., Bronfman, L., \& Knude, J.\ 2007, \aap, 463, 1029 
\bibitem[Kainulainen et al.(2009a)]{kai09pipe} Kainulainen, J., Lada, C.~J., Rathborne, J.~M., \& Alves, J.~F.\ 2009, \aap, 497, 399 
\bibitem[Kainulainen et al.(2009b)]{kai09} Kainulainen, J., Beuther, H., Henning, T., \& Plume, R.\ 2009, \aap, 508, L35 
\bibitem[Kenyon et al.(2008)]{ken08} Kenyon, S.~J., G{\'o}mez, M., \& Whitney, B.~A.\ 2008, Handbook of Star Forming Regions, Volume I, 405
\bibitem[Lada et al.(1994)]{lad94} Lada, C.~J., Lada, E.~A., Clemens, D.~P., \& Bally, J.\ 1994, \apj, 429, 694
\bibitem[Lada et al.(2008)]{lad08} Lada, C.~J., Muench, A.~A., Rathborne, J., Alves, J.~F., \& Lombardi, M.\ 2008, \apj, 672, 410 
\bibitem[Larson(1981)]{lar81} Larson, R.~B.\ 1981, \mnras, 194, 809 
\bibitem[Lee \& Myers(1999)]{lee99} Lee, C.~W., \& Myers, P.~C.\ 1999, \apjs, 123, 233 
\bibitem[Loinard et al.(2007)]{loi07} Loinard, L., Torres, R.~M., Mioduszewski, A.~J., Rodr{\'{\i}}guez, L.~F., Gonz{\'a}lez-L{\'o}pezlira, R.~A., Lachaume, R., V{\'a}zquez, V., \& Gonz{\'a}lez, E.\ 2007, \apj, 671, 546 
\bibitem[Lombardi \& Alves(2001)]{lom01} Lombardi, M., \& Alves, J.\ 2001, \aap, 377, 1023 
\bibitem[Lombardi et al.(2006)]{lom06} Lombardi, M., Alves, J., \& Lada, C.~J.\ 2006, \aap, 454, 781
\bibitem[Lombardi(2009)]{lom09} Lombardi, M.\ 2009, \aap, 493, 735
\bibitem[Lombardi et al.(2010)]{lom10} Lombardi, M., Lada, C.~J., \& Alves, J.\ 2010, \aap, 512, A67 
\bibitem[Luhman et al.(2003)]{luh03} Luhman, K.~L., Brice{\~n}o, C., Stauffer, J.~R., Hartmann, L., Barrado y Navascu{\'e}s, D., \& Caldwell, N.\ 2003, \apj, 590, 348 
\bibitem[Luhman et al.(2009)]{luh09} Luhman, K.~L., Mamajek, E.~E., Allen, P.~R., \& Cruz, K.~L.\ 2009, \apj, 703, 399
\bibitem[Lynds(1962)]{lyn62} Lynds, B.~T.\ 1962, \apjs, 7, 1 
\bibitem[McGlynn et al.(1998)]{mcg98} McGlynn, T., Scollick, K., \& White, N.\ 1998, New Horizons from Multi-Wavelength Sky Surveys, 179, 465
\bibitem[Men'shchikov et al.(2010)]{men10} Men'shchikov, A., et al.\ 2010, \aap, 518, L103 
\bibitem[Mizuno et al.(1995)]{miz95} Mizuno, A., Onishi, T., Yonekura, Y., Nagahama, T., Ogawa, H., \& Fukui, Y.\ 1995, \apjl, 445, L161 
\bibitem[Molinari et al.(2010)]{mol10} Molinari, S., et al.\ 2010, \aap, 518, L100 
\bibitem[Motte et al.(1998)]{mot98} Motte, F., Andre, P., \& Neri, R.\ 1998, \aap, 336, 150 
\bibitem[Myers et al.(1991)]{mye91} Myers, P.~C., Fuller, G.~A., Goodman, A.~A., \& Benson, P.~J.\ 1991, \apj, 376, 561
\bibitem[Myers(2009)]{mye09} Myers, P.~C.\ 2009, \apj, 700, 1609
\bibitem[Narayanan et al.(2008)]{nar08} Narayanan, G., Heyer, M.~H., Brunt, C., Goldsmith, P.~F., Snell, R., \& Li, D.\ 2008, \apjs, 177, 341 
\bibitem[Ochsenbein et al.(2000)]{vizier} Ochsenbein, F., Bauer, P., \& Marcout, J.\ 2000, \aaps, 143, 23 
\bibitem[Onishi et al.(1998)]{oni98} Onishi, T., Mizuno, A., Kawamura, A., Ogawa, H., \& Fukui, Y.\ 1998, \apj, 502, 296
\bibitem[Onishi et al.(2002)]{oni02} Onishi, T., Mizuno, A., Kawamura, A., Tachihara, K., \& Fukui, Y.\ 2002, \apj, 575, 950
\bibitem[Ostriker(1964)]{ost64} Ostriker, J.\ 1964, \apj, 140, 1056
\bibitem[Palla \& Stahler(2002)]{pal02} Palla, F., \& Stahler, S.~W.\ 2002, \apj, 581, 1194 
\bibitem[Pilbratt et al.(2010)]{pil10} Pilbratt, G.~L., et al.\ 2010, \aap, 518, L1
\bibitem[Pineda et al.(2008)]{pin08} Pineda, J.~E., Caselli, P., \& Goodman, A.~A.\ 2008, \apj, 679, 481 
\bibitem[Pineda et al.(2009)]{pin09} Pineda, J.~E., Rosolowsky, E.~W., \& Goodman, A.~A.\ 2009, \apjl, 699, L134
\bibitem[Quanz et al.(2010)]{qua10} Quanz, S.~P., Goldman, B., Henning, T., Brandner, W., Burrows, A., \& Hofstetter, L.~W.\ 2010, \apj, 708, 770 
\bibitem[Rathborne et al.(2009)]{rat09} Rathborne, J.~M., Lada, C.~J., Muench, A.~A., Alves, J.~F., Kainulainen, J., \& Lombardi, M.\ 2009, \apj, 699, 742
\bibitem[Rebull et al.(2010)]{reb10} Rebull, L.~M., et al.\ 2010, \apjs, 186, 259 
\bibitem[Rieke \& Lebofsky(1985)]{rie85} Rieke, G.~H., \& Lebofsky, M.~J.\ 1985, \apj, 288, 618 
\bibitem[Rom{\'a}n-Z{\'u}{\~n}iga et al.(2009)]{rom09} Rom{\'a}n-Z{\'u}{\~n}iga, C.~G., Lada, C.~J., \& Alves, J.~F.\ 2009, \apj, 704, 183 
\bibitem[Rosolowsky et al.(2008)]{ros08} Rosolowsky, E.~W., Pineda, J.~E., Kauffmann, J., \& Goodman, A.~A.\ 2008, \apj, 679, 1338 
\bibitem[Ryden(1996)]{ryd96} Ryden, B.~S.\ 1996, \apj, 471, 822 
\bibitem[Sadavoy et al.(2010)]{sad10} Sadavoy, S.~I., et al.\ 2010, \apj, 710, 1247 
\bibitem[Schneider \& Elmegreen(1979)]{sch79} Schneider, S., \& Elmegreen, B.~G.\ 1979, \apjs, 41, 87 
\bibitem[Siess et al.(2000)]{sie00} Siess, L., Dufour, E., \& Forestini, M.\ 2000, \aap, 358, 593 
\bibitem[Skrutskie et al.(2006)]{2mass} Skrutskie, M.~F., et al.\ 2006, \aj, 131, 1163
\bibitem[Stutzki \& Guesten(1990)]{stu90} Stutzki, J., \& Guesten, R.\ 1990, \apj, 356, 513 
\bibitem[Terebey et al.(2010)]{ter10} Terebey, S., Fich, M., Noriega-Crespo, A., Padgett, D., \& Taurus Spitzer Legacy Team 2010, Bulletin of the American Astronomical Society, 41, 559 
\bibitem[Torres et al.(2007)]{tor07} Torres, R.~M., Loinard, L., Mioduszewski, A.~J., \& Rodr{\'{\i}}guez, L.~F.\ 2007, \apj, 671, 1813 
\bibitem[Torres et al.(2009)]{tor09} Torres, R.~M., Loinard, L., Mioduszewski, A.~J., \& Rodr{\'{\i}}guez, L.~F.\ 2009, \apj, 698, 242
\bibitem[Vasyunina et al.(2009)]{vas09} Vasyunina, T., Linz, H., Henning, T., Stecklum, B., Klose, S., \& Nyman, L.-{\AA}.\ 2009, \aap, 499, 149 
\bibitem[Williams et al.(1994)]{clumpfind} Williams, J.~P., de Geus, E.~J., \& Blitz, L.\ 1994, \apj, 428, 693
\end{thebibliography}
\end{document}